\documentclass[prl,aps,twocolumn,showpacs]{revtex4}%
\usepackage{graphicx,dcolumn,amsmath,color,longtable}
\usepackage{amsmath}
\usepackage{amsfonts}
\usepackage{amssymb}
\usepackage{graphicx}%
\newcommand{\bk}{{\mathbf k}}
\newcommand{\bp}{{\mathbf p}}
\newcommand{\br}{{\mathbf r}}
\newcommand{\be}{\begin{equation}}
\newcommand{\ee}{\end{equation}}
\newcommand{\bea}{\begin{eqnarray}}
\newcommand{\eea}{\end{eqnarray}}

\begin{document}
\title{Theory of anisotropic Rashba splitting of surface states}
\author{E. Simon$^{1,2}$}
\author{A. Szilva$^{3}$}
\author{B. Ujfalussy$^{1}$}
\author{B. Lazarovits$^{1,3}$}
\author{G. Zarand$^{3}$}
\author{L. Szunyogh$^{3}$}
\email{szunyogh@phy.bme.hu}
\affiliation{
$^{1}$Hungarian Academy of Sciences, Institute for Solid State Physics and Optics, H-1525 Budapest, PO Box 49 H-1525 Hungary \\
$^{2}$L\'or\'and E\"otv\"os University, Department of Physics, H-1518 Budapest POB 32, Hungary \\
$^{3}$Department of Theoretical Physics, Budapest University of Technology and
Economics, Budafoki \'ut 8., H-1111 Budapest, Hungary}
\date{\today}

\begin{abstract}
We investigate the surface Rashba effect for a surface of reduced in-plane
symmetry. Formulating a \textbf{k}$\cdot$\textbf{p} perturbation 
theory, we show that the Rashba splitting is anisotropic, in agreement with 
symmetry-based considerations. We show that 
the  anisotropic Rashba splitting  is due to the admixture of 
bulk states of different symmetry to the surface state, and 
it cannot be explained within the standard theoretical 
picture supposing just a normal-to-surface variation of the crystal potential. 
Performing  relativistic \emph{ab
initio} calculations we find a remarkably large Rashba anisotropy  
for an unreconstructed Au(110) surface that is in the experimentally accessible range.
\end{abstract}
\maketitle

Metallic surfaces often exhibit Shockley-type surface states 
located in a relative band gap of the bulk band structure, and
 forming a two-dimensional electron
gas. 
One of the most intriguing manifestation of spin-orbit coupling (SOC) at
surfaces is the splitting of these surface states, 
known as Rashba splitting~\cite{Rashba-spss60,BR-jetp84}.
Such Rashba splitting was observed via
photoemission by LaShell \emph{et al.}~\cite{LMJ-prl96} for the $L$-gap
surface state at Au(111) and explained theoretically in terms of a
tight-binding model~\cite{PH-ss00} and \emph{ab
initio} electronic structure calculations~\cite{NRH+prb01,HEB-prb03}, but 
several studies of the Rashba splitting were published in 
recent years on Bi(111) and Bi/Ag(111)~\cite{KGB+prl04,PTHB-prb07,AHE+prl07},
as well as on Bi$_{x}$Pb$_{1-x}$%
/Ag(111), where  atomic Bi 
$p$-orbitals lead to a more pronounced spin-orbit splitting~\cite{BBC-prb07,APM+prb08,HKS+prb08,DML+prl08}.

Describing and controlling the 
Rashba splitting of surface states is  crucial 
for spintronics applications. The famous Datta-Das transistor relies 
on the electric tuning of the Rashba splitting~\cite{DattaDas} and   
the Rashba splitting is responsible for the spin Hall effect in two 
dimensions~\cite{spinHall} and the anomalous Hall effect~\cite{AHE} 
as well.

The simplest way to understand the origin of the Rashba effect is 
to take nearly free electrons, confined by a crystal potential, 
$V({\mathbf r}) = V(z)$, and having a plane-wave-like wave function, 
$\psi_{s,\mathbf{k}}\left(  \mathbf{r}\right)  =
e^{i\mathbf{kr}}
\phi\left(z\right)  \chi_{s}$,  with $\chi_{s}$ some spinor 
eigenfunctions, and $\mathbf k$ the momentum parallel to the surface.  
The crystal potential $V(z)$ obviously produces an electric field, 
${\bf E}$, perpendicular to the surface, which, in the presence of spin-orbit 
interaction
leads to the following spin-orbit term in the effective Hamiltonian, 
\begin{equation}
H_{R}\left(  \mathbf{k}\right)  
=\alpha_{R}\left(  k_{x}\sigma
_{y}-k_{y}\sigma_{x}\right)  \;, 
\label{eq:RH-simple}%
\end{equation}
called Rashba-Hamiltonian. In Eq. (\ref{eq:RH-simple}), ${\sigma}_i$ 
denote the Pauli
matrices and $\alpha_{R}=\frac{\hbar^{2}}{4m^{2}c^{2}}\int d^{3}r\,|\phi\left(  z\right)|^2  {\partial_z V(\mathbf{r})}$ 
is the so-called Rashba parameter. The eigenvalue problem can then
easily be solved, resulting in a splitting of the spin-degeneracy
of the surface states,
$
\varepsilon_{\pm}\left(  \mathbf{k}\right)  
=\frac{\hbar^{2}}{2m^{\ast}}\mathbf{k}%
^{2}\pm\alpha_{R}\,\left\vert \mathbf{k}\right\vert,
$
with $m^{\ast}$ the effective mass of the surface 
electrons~\cite{PH-ss00,HEB-prb03}. 
Clearly, the above dispersion is
isotropic in k-space, hence we term it as isotropic 
Rashba splitting. 

Although real systems cannot be described in 
terms of free electrons, and for quantitative estimates of $\alpha_R$
 the atomic structure of the potential 
needs be taken into account~\cite{PTHB-prb07}, the 
structure of the Rashba interaction, Eq.~\eqref{eq:RH-simple}, is 
very robust for surfaces of high point-group 
symmetry such as $C_{3v}$ or $C_{4v}$~\cite{OS-jpcm09}.

\begin{figure}[t]
\begin{center}
\includegraphics[height=5cm]{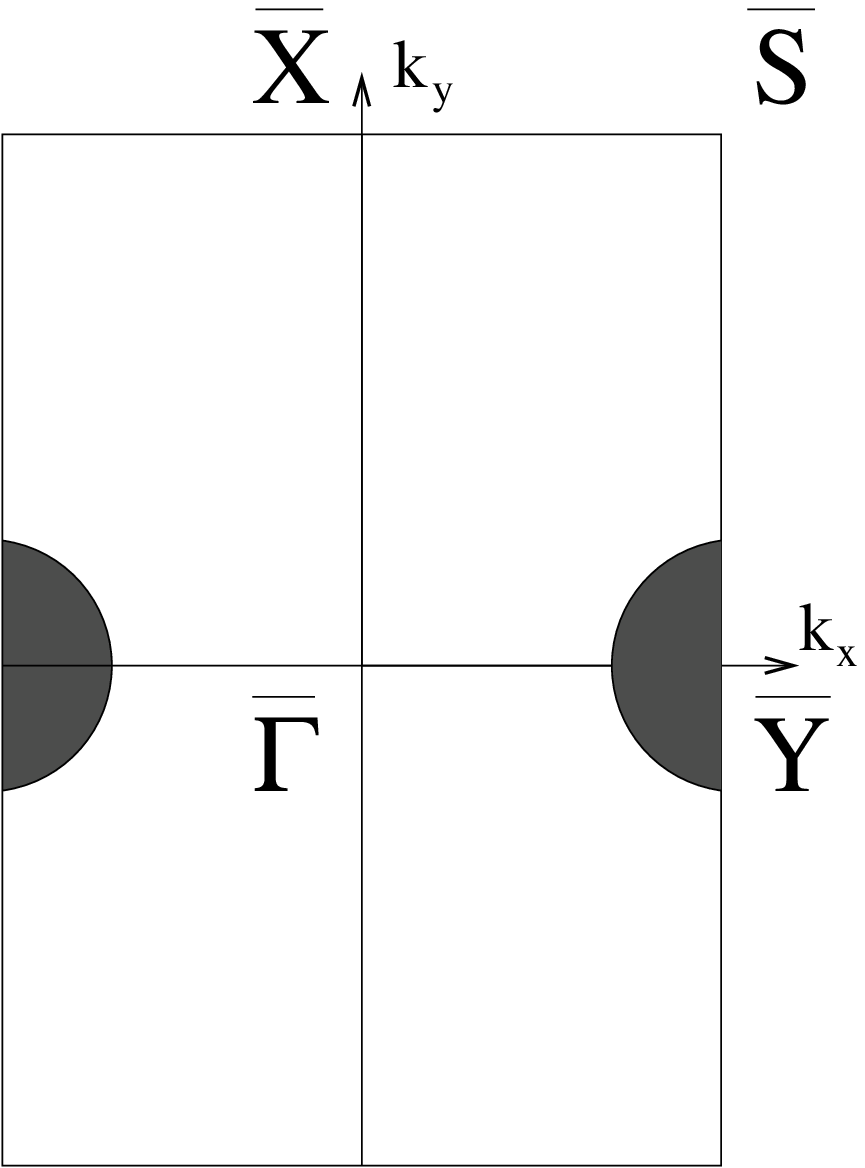}\hfill\includegraphics[height=4.5cm]{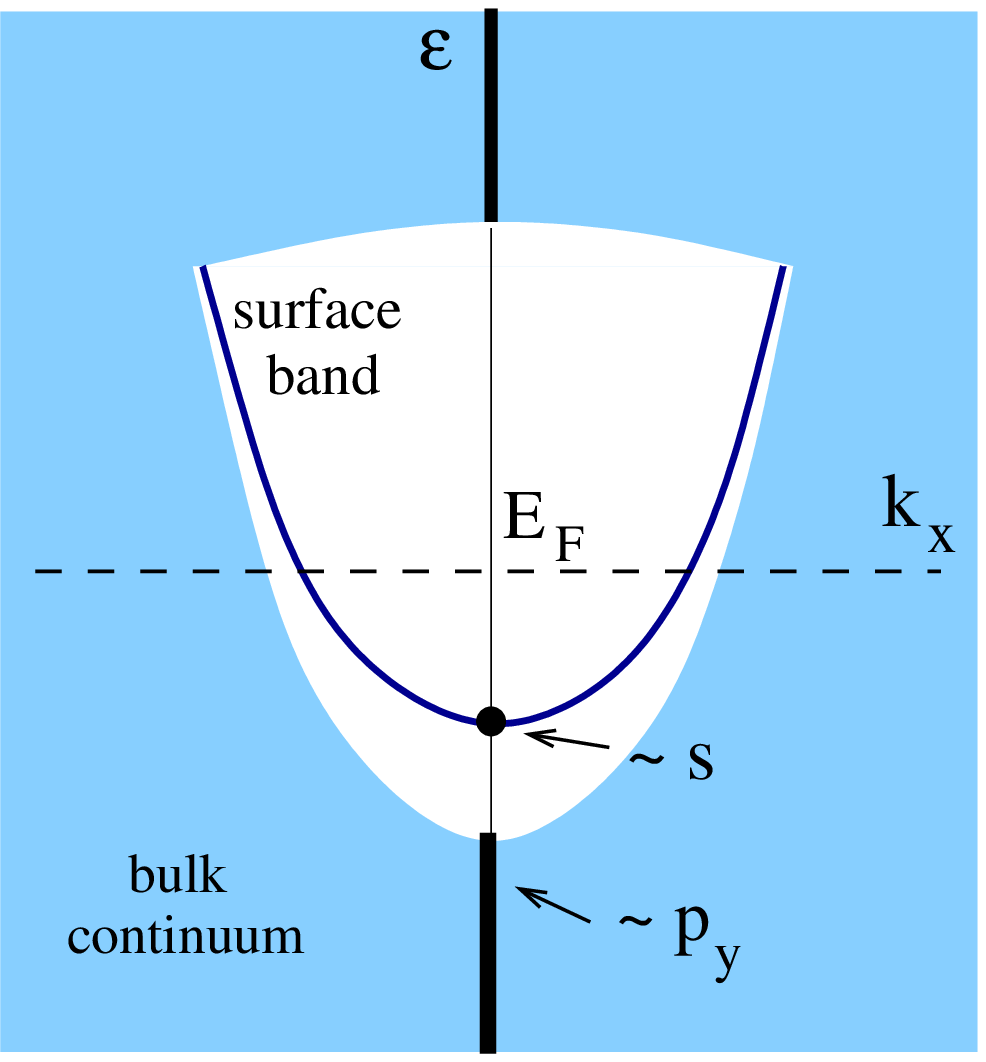}
\end{center}
\par
\vskip -12pt\caption{(Color online) Left: Sketch of the fcc(110) 
Surface Brillouin Zone. The dark
area denotes the projection of the $L$-gap of bulk Au. Right: Structure 
of the surface energy spectrum in the absence of SO interaction, along the line 
$\bk=(k_x,0)$. Surface states in the relative gap with 
$\bk\ne0$ can be built up from states 
indicated by the thick black lines and the black circle at $\bk=0$.
Note that $\bk=0$ corresponds to the ${\overline{\mbox{Y}}}$ point of the 
Brillouin zone,see Eq.~(\ref{eq:psik}).}
\label{fig:BZ}%
\end{figure}

The situation is, however, quite different for surfaces 
(or points in the surface Brillouin zone) of  reduced symmetry.  
Such  Shockley-type surface
states emerge, e.g., around the $\overline{\text{Y}}$ point of 
the Surface Brillouin Zone of
unreconstructed and (2$\times$1) reconstructed Au(110) surfaces, 
as revealed by  recent high-resolution photoelectron spectroscopy
experiments~\cite{NHF+prb08}.  
In this case, the $C_{2v}$
point-group symmetry of the system not only implies the asymmetry of the
effective mass, $m_{x}^{\ast}\neq m_{y}^{\ast}$ (for the 
crystal axes see Fig.\ref{fig:BZ}) but, in leading order in $\mathbf k$, 
representation theory also   predicts the following simple form 
of the effective Hamiltonian~\cite{OS-jpcm09},  
\begin{equation}
H\left(  \mathbf{k}\right)  =\varepsilon_{0}+\frac{\hbar^{2}k_{x}^{2}}%
{2m_{x}^{\ast}}+\frac{\hbar^{2}k_{y}^{2}}{2m_{y}^{\ast}}+\alpha_{R,x}%
\,k_{x}\sigma_{y} - \alpha_{R,y}\,k_{y}\sigma_{x}
\;.
\label{eq:Heffk}%
\end{equation}
The above expression can easily be justified by  simple symmetry 
analysis, just by  noticing  that $\sigma_y$ and $-\sigma_x$ 
transform as $p_x$ and $p_y$ under the operations of the double 
groups of $C_{2v}$ and $C_{4v}$. 
From this observation it also follows 
that in case of $C_{4v}$ point-group symmetry 
$\alpha_{R,x}=\alpha_{R,y}$
must be satisfied, and the Hamiltonian (\ref{eq:RH-simple}) is 
recovered.~\cite{remark}

Although the above form of the Rashba interaction has been predicted in 
Ref.~\cite{OS-jpcm09}, no microscopic theory has been constructed 
so far to support it. While previous \emph{ab
initio} calculations~\cite{NHF+prb08,NKSO+jpcm09} did find
a Rashba splitting of the Au(110) surface state, they focused only on the 
dispersion along the $\overline{\Gamma}\overline{\text{Y}}$ direction, and therefore the 
anisotropy of the Rashba term remained unnoticed. 
In the present paper, 
we provide such a microscopic analysis
for an Au(110) surface with $C_{2v}$ point-group
symmetry. First, constructing a 
$\mathbf{k\cdot p}$ perturbation theory for the surface states 
we show that the 
above anisotropic Rashba structure appears naturally, and 
is due to the finite momentum 
mixing of the bulk $p$ states to the surface-state. 
We also perform \emph{ab initio} calculations of
the Rashba-split surface state of an unreconstructed Au(110) surface and
confirm  with a high numerical accuracy
that there is a large anisotropy in $k$-space, 
$\alpha_{R,x}\sim 5 \, \alpha_{R,y}$, in agreement with Eq.~\eqref{eq:Heffk}.
The predicted  anisotropic Rashba splittings turn out to be 
within the range of  experimental accuracy.  
  

Bloch-states of Au(110) can be characterized by a surface 
momentum, and can thus be written as 
\begin{equation}
\psi_{\mathbf{Q+k}}\left(  \mathbf{r}\right)  =e^{i\mathbf{kr}}\phi
_{\mathbf{Q},\mathbf{k}}\left(  \mathbf{r}\right)  \;, 
\label{eq:psik}
\end{equation}
with the momentum $\mathbf k$ measured with respect to the 
momentum $\mathbf{Q}$  associated with the $\overline{\text{Y}}$
point of the Surface Brillouin Zone. Here the functions 
$\phi
_{\mathbf{Q},\mathbf{k}}\left(  \mathbf{r}\right)$
are lattice-antiperiodic in the $x$ direction, while they are 
lattice periodic in the $y$ direction of the (110) plane, see Fig. 1.
For any given momentum, 
$\bk$, there exist an infinite number (continuum) of eigenstates, 
 the energy of which ($\varepsilon_\bk$) is determined by the condition that the states 
$\psi_{\mathbf{Q+k}}$ be eigenstates of the Hamiltonian, 
$H = \frac {\bp^2}{2m} + V(\br) + H_{SO}$, with $H_{SO}$ denoting 
the spin-orbit  coupling,
\begin{equation}
H_{SO}\left(  \mathbf{r}\right)  =\frac{\hbar}{4m^{2}c^{2}}\left(  \nabla
V\left(  \mathbf{r}\right)  \times\mathbf{p}\right)  \mathbf{\sigma\;.}%
\end{equation}
As a consequence, the functions $\phi
_{\mathbf{Q},\mathbf{k}}$ must satisfy the equation, 
\textbf{\ }%
\begin{align}
&  \left(  \frac{\mathbf{p}^{2}}{2m}+V\left(  \mathbf{r}\right)  +\frac
{\hbar^{2}\mathbf{k}^{2}}{2m}+\frac{1}{m}\mathbf{k\cdot p}+{\tilde H}_{SO}\left(
\mathbf{k},\mathbf{r}\right)  \right)  \phi_{\mathbf{Q},\mathbf{k}}\left(
\mathbf{r}\right) \nonumber\\
&  \quad=\varepsilon_{\mathbf{k}}\phi_{\mathbf{Q},\mathbf{k}}\left(
\mathbf{r}\right)  \;, \label{eq:PS-SOC}%
\end{align}
with ${\tilde H}_{SO}\left(  \mathbf{k},\mathbf{r}\right)  $
being the effective SO coupling, 
\begin{equation}
{\tilde H}_{SO}\left(  \mathbf{k},\mathbf{r}\right)  =
H_{SO}\left(  \mathbf{r}\right)  \; 
+ 
\frac{\hbar}{4m^{2}c^{2}%
}\left(  \nabla V\left(  \mathbf{r}\right)  \times\hbar\mathbf{k}\right)
\mathbf{\sigma
\;.}
\label{eq:tildeH_SO}
\end{equation}
 Similar to 
Bloch wave functions, for any fixed momentum, $\bk$, 
(and for any value of $\tilde H_{SO}$) 
the functions  $\phi_{\mathbf{Q},\mathbf{k}}$  form a 
complete set for functions having the previously-mentioned periodicity 
property. In the spirit of $\bk\cdot\bp$ perturbation theory, we can thus 
take the complete set of $\bk = 0$ and $\tilde H_{SO}=0$ solutions, 
satisfying 
\begin{equation}
\left(  \frac{\mathbf{p}^{2}}{2m}+V\left(  \mathbf{r}\right)  \right)
\phi_{i, n_i}\left(  \mathbf{r}\right)  =\varepsilon
_{i,n_i}\phi_{i,n_i}\left(  \mathbf{r}%
\right)  \;,
\end{equation}
and expand  $\phi_{\mathbf{Q},\mathbf{k}}$ in terms of these. 
Here we classified the solutions according to the four one-dimensional
irreducible representations of the  $C_{2v}$ symmetry associated with the 
point $\overline{\text{Y}}$, $i \in \{1,x,y,xy\}$, 
and labeled solutions of a given symmetry by $n_i$.
As shown in Fig.~\ref{fig:BZ}, the spectrum contains a discrete 
surface state of
$s$-symmetry and the projected bulk continuum forming the gap. Let us denote
the $\bk=0$ surface state by $\phi_{0}$, and its  eigenenergy 
by $\varepsilon_{0}$. 
Then states with $\bk\ne 0$ 
but with $\tilde H_{SO}\equiv0$ can be expressed in terms of the 
states $\phi_{i,n_i}$ by performing 
second order perturbation theory in $\bk$, which amounts in  a surface state
\be
\left\vert \phi_{\mathbf{k}}^{0}\right\rangle  = \left\vert \phi_{0}%
\right\rangle +\frac{1}{m}\sum_{{i,n_i}\left(  \neq0\right)  }
\frac{\left\vert \phi_{i,n_i}\right\rangle \left\langle
\phi_{i,n_i}\left\vert \mathbf{k\cdot p}\right\vert
\phi_{0}\right\rangle }{\varepsilon_{0}-\varepsilon_{i,n_i}}\;,
\label{eq:wfk0}%
\ee
with approximate dispersion
\bea
\varepsilon_{\mathbf{k}}^{0}  &  = &
\varepsilon_{0}+\frac{\hbar^{2}k_{x}^{2}}{2m_{x}^{\ast}}+\frac{\hbar
^{2}k_{y}^{2}}{2m_{y}^{\ast}}\;,
\label{eq:m*}
\\
\frac{1}{{m_i}^{\ast}}&=&\frac{1}{m}+\frac{2}{m^{2}}\sum_{n_i} 
\frac{\left\vert \left\langle \phi_{i,n_i} \left\vert p_{i} 
\right\vert \phi_{0} \right\rangle \right\vert ^{2} }
{\varepsilon_{0}-\varepsilon_{i,n_i}} \quad\left(  i=x,y\right)  \;. 
\eea
The index 0 in $\varepsilon_\bk^0$ and 
$\left\vert \phi_\bk^0\right\rangle$ is meant to remind us to the
absence of SO interaction.    

To obtain the surface states, $\left\vert \phi_\bk\right\rangle$, 
we then carry out    first-order perturbation theory with the SOC
operator, $\tilde H_{SO}$, using  the states 
$\left\vert \phi_\bk^0\right\rangle$ as a starting point. 
Keeping just contributions linear 
in $\mathbf{k}$ we get two terms to the effective Rashba 
Hamiltonian. The second term in Eq.~\eqref{eq:tildeH_SO} gives rise
to the usual isotropic Rashba model, 
\begin{equation}
H_{R}^{iso}\left(  \mathbf{k}\right) 
={\alpha}_{R} \left( \mathbf{e}_{z} \times\mathbf{k}\right)\cdot  \mathbf{\sigma
\;,}\label{eq:HRiso}%
\end{equation}
with 
$
\alpha_{R}=
\frac {\hbar^{2} }{4m^{2}c^{2}}
\left\langle \phi_{0}\right\vert {\partial_z V}/{\partial
z}\left\vert \phi_{0}\right\rangle
$.
 The term $H_{SO}$ in Eq.~\eqref{eq:tildeH_SO}, however,  
  gives also a finite contribution due to the admixture of 
$p_{x,y}$ states from the continuum and, in fact, this is 
  precisely the term that leads to an anisotropic Rashba coupling,
\begin{align}
&  H_{R}^{\rm anis}\left(  \mathbf{k}\right)  = \\
&  =\frac{1}{m}\sum_{i=x,y}k_{i}\sum_{n_i}\frac{\left\langle
\phi_{i,n_i}\left\vert p_i\right\vert \phi
_{0}\right\rangle \left(  \left\langle \phi_{0}\right\vert \mathbf{a}%
\left\vert \phi_{i,n_i}\right\rangle \,\mathbf{\sigma
}\right)  +h.c.}{\varepsilon_{0}-\varepsilon_{i,n_i}}\;,
\nonumber
\end{align}
where we defined the (axial)vector operator related to SOC, $\mathbf{a}%
=\frac{\hbar}{4m^{2}c^{2}}\left(  \nabla V\left(  \mathbf{r}\right)
\times\mathbf{p}\right)  $. Using the symmetry of the unperturbed
wave functions, a particularly simple form of the above anisotropic Rashba
Hamiltonian can be obtained,%
\be
H_{R}^{anis}\left(  \mathbf{k}\right)  =\lambda_{x}k_{x}\sigma_{y}+\lambda
_{y}k_{y}\sigma_{x}\;,%
\label{eq:lamxy} 
\ee
with the coefficient $\lambda_x$ expressed as
\be
\lambda_{x}   =\frac{2}{m}\sum_{n_{x}}\frac{\operatorname{Re}\left(
\left\langle \phi_{x\mathbf{,}n_{x}}\left\vert p_{x}\right\vert \phi
_{0}\right\rangle \left\langle \phi_{0}\right\vert a_{y}\left\vert
\phi_{x\mathbf{,}n_{x}}\right\rangle \right)  }{\varepsilon_{0}-\varepsilon
_{x,n_{x}}}\;, 
\label{eq:lambda}
\ee
and $\lambda_y$ given by a similar expression. The structure of this
term is identical  to the one obtained by symmetry analysis. 
Decreasing the in-plane
asymmetry of the potential (e.g., considering $C_{4v}$ point group symmetry),
the relationship, $\lambda_{x}=-\lambda_{y}$, is satisfied, {\em i.e.}, the Rashba
splitting becomes isotropic. Nevertheless, even in this case, the anisotropic
Rashba term, Eq. (\ref{eq:lamxy}), arising from the mixing of the
surface state with bulk states, also contributes to the Rashba splitting. 
Thus we find that, up to second order in \textbf{k}, 
 surface states of Au(110) are described by Eq.~\eqref{eq:Heffk}, with 
$\alpha_{R,x}=\alpha_{R}+\lambda_{x}$ and 
$\alpha_{R,y}=-\alpha_{R}+\lambda_{y}$. 

\begin{figure}[th]
\begin{center}
\includegraphics[width=8.0cm,bb=0 0 410 280,clip]{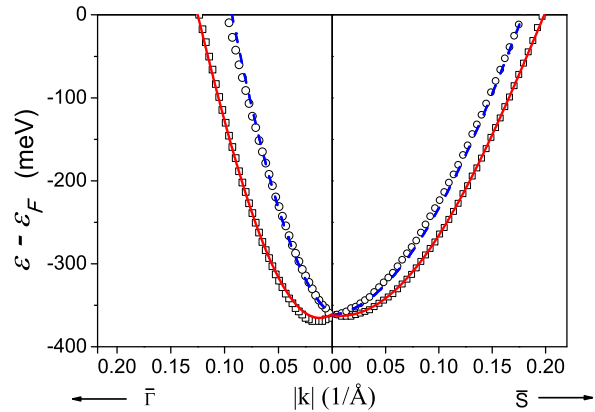}
\end{center}
\vskip -12pt
\caption{(Color online) Dispersion relations of the
Au(110) surface states at the $\overline{\mbox Y}$ point ($|k|=0$)
along the $\overline{\mbox{Y}\Gamma}$ and 
the $\overline{\mbox{YS}}$ 
directions. Symbols refer to the calculated data, solid and dashed lines 
to the fitted curves for $\varepsilon_-({\bf k})$ and $\varepsilon_+({\bf k})$, respectively. 
}%
\label{fig:disp}
\end{figure}

To obtain a quantitative estimate of the parameters $\alpha_{R,x/y}$ and the 
induced Rashba splittings,  we performed
calculations of the surface states of unreconstructed Au(110) surface near the
$\overline{\text{Y}}$ point of the Surface Brillouin zone, using
the relativistic Screened Korringa-Kohn-Rostoker (KKR) method.
Details of this method  are described in
Refs.~\cite{RSKKR,SKKR-book}. The computed dispersion relations along the $\overline{\Gamma\mbox{Y}}$ ($x$) and
the $\overline{\mbox{YS}}$ ($y$) directions
 are
plotted in Fig.~\ref{fig:disp}. The maximum binding energy,
$\varepsilon_{0}\simeq 370$ meV, is by about 200~meV less than the
measured value~\cite{NHF+prb08} and other theoretical
values~\cite{NHF+prb08,NKSO+jpcm09}. This deviation is mostly caused 
by the atomic sphere approximation (ASA) and 
the angular momentum cut-off, $\ell_{max}=2$, which resulted in 
some error for the determination of the Fermi level and the vacuum potential. 

The nearly free electron-like, parabolic shape of the dispersion 
as well as the Rashba splitting being remarkably different along the two
directions is obvious from Fig.~\ref{fig:disp}, and a detailed analysis 
confirms this impression: The numerical results are very well fitted 
by the dispersions
$
\varepsilon_{\pm}\left(  \mathbf{k}\right)  =\varepsilon_{0}+\frac{\hbar
^{2}k_{x}^{2}}{2m_{x}^{\ast}}+\frac{\hbar^{2}k_{y}^{2}}{2m_{y}^{\ast}}\pm
\sqrt{\alpha_{R,x}^{2}k_{x}^{2}+\alpha_{R,y}^{2}k_{y}^{2}}\;,
$
obtained by diagonalizing the approximate Hamiltonian, 
   Eq. (\ref{eq:Heffk}), with the fitting parameters,  
$m_x^*=0.11 \, m$, 
$m_y^*=0.32 \, m$,
$\alpha_{R,x} = 0.8$ eV~\AA,  and $\alpha_{R,y} = 0.17$ eV~\AA.
The obtained effective mass along the $\overline{\mbox{YS}}$ direction 
is in satisfactory agreement with the measured value, 
$m_y=0.25 \, m$.~\cite{NHF+prb08}  
The effective mass along $\overline {\Gamma\mbox{Y}}$, $m_x$ is only about 
one  third of $m_y$, which is the consequence that the 
states at the lower bulk band edge are mainly of $p_z$ and $p_x$ character
(see  Eq.~(\ref{eq:m*})). Note that the energy separation of 
the surface state at the 
$\overline{\mbox Y}$ point is 0.8 and 3.4 eV with respect to the lower and 
upper bulk band edges, respectively, implying a strong admixture
of ``electron'' states from the continuum below the surface state.

One of the most astonishing results of these numerical 
calculations is the  remarkably large anisotropy of the
Rashba parameters, $\alpha_{R,x} \sim 5 \,\alpha_{R,y}$. In view of 
Eqs.~\eqref{eq:lambda}, this observation can also be
explained with the absence of $p_y$ states at the lower bulk 
band edge. This result also correlates with the results of 
the effective mass: the smaller value of $m_x$ indicates a stronger admixture 
of $p_x$ states, also responsible for the stronger renormalization 
of $\alpha_x$.
We remark that $\alpha_{R,x}$ is even larger than 
$\alpha_{R}$ we calculated for the $L$-gap state of Au(111), 0.57 eV~\AA .
This latter value is though considerably larger than the experimental one,
0.4 eV~\AA ~\cite{HHO+jpcm04}, which  correlates with  
the theoretically computed effective mass,
$m^\ast \sim 0.19 \, m_e$, being too small
as compared to the experimentally observed value, $m^\ast \sim 0.25\, m_e$.
The computed Fermi wave numbers, $k_F$= 0.160  and 0.189 \AA$^{-1}$,
on the other hand, 
are almost in perfect agreement with the measured values~\cite{HHO+jpcm04}.
Nevertheless, based upon the discrepancy regarding the value of the 
effective masses, we expect that our theoretical calculations for 
Au(110) somewhat overestimate  the Rashba parameters, $\alpha_{R,x/y}$.  

\begin{figure}[th]
\begin{center}
\includegraphics[width=7.0cm,bb=10 10 225 220]{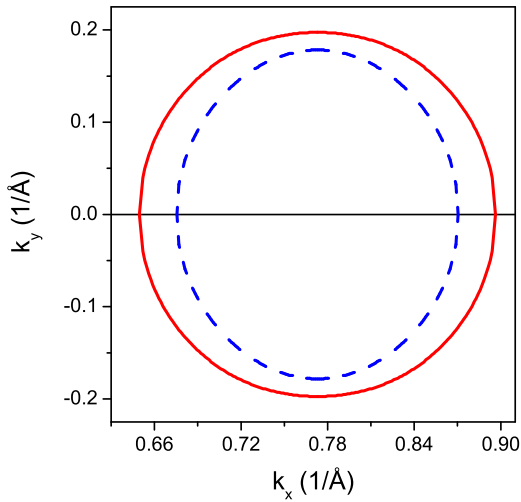}
\end{center}
\par
\vskip -12pt\caption{(Color online) Rashba splitting of the Au(110) surface
state at the Fermi level. Solid and dashed lines refer to the bands,
$\varepsilon_+({\bf k})$ and $\varepsilon_-({\bf k})$, respectively. 
}
\label{fig:ASS}%
\end{figure}
\begin{figure}[th]
\begin{center}
\includegraphics[width=7.0cm,bb=10 10 360 300]{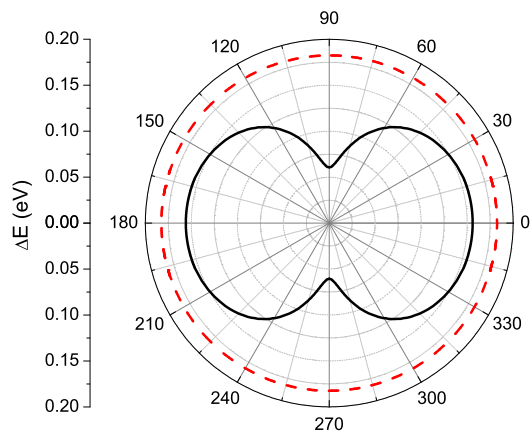}
\end{center}
\par
\vskip -12pt\caption{
(Color online) Energy differences, $\Delta \varepsilon({\bf k}) = 
\varepsilon_+({\bf k}) - \varepsilon_-({\bf k})$,
 for the Rashba-split surface state of
Au(110) (solid line) and Au(111) (dashed line)  
as a function of the polar angle $\varphi=
\arctan \frac{k_y}{k_x}$, shown in units of degree around the graph.
The magnitude of {\bf k}
was fixed to satisfy $\varepsilon_+({\bf k})=\varepsilon_F$.
The energy scale is indicated by the axis on the left.}
\label{fig:polar}%
\end{figure}

The anisotropic Rashba coupling together with the anisotropic 
effective mass gives rise to two Rashba-split Fermi surfaces 
for the surface states, as shown in Fig.~\ref{fig:ASS}. 
The Rashba splitting along $\overline{\mbox{YS}}$, $\Delta k_y 
\simeq$  0.017 \AA$^{-1}$ is in the order of the
experimental resolution (0.01~\AA$^{-1}$)~\cite{NHF+prb08},
but the value $\Delta k_x \simeq$ 0.026 \AA$^{-1}$ 
implies that the Rashba splitting should be detectable experimentally
along the $\overline {\Gamma\mbox{Y}}$ direction. 

In Fig.~\ref{fig:polar}  we also show the polar plot of 
 the energy splitting, $\Delta \varepsilon({\bf k}) =
\varepsilon_+({\bf k}) - \varepsilon_-({\bf k})$,
 between the two bands at the inner Fermi 
surface ($\varepsilon_+({\bf k})=\varepsilon_F$). 
As a comparison, the same quantity is displayed for the surface state of
Au(111). Supporting the above implication, $\Delta \varepsilon({\bf k})$
along $\overline{\mbox{Y}\Gamma}$ for Au(110) 
is almost as large as the (isotropic)
energy splitting in case of Au(111).
Furthermore, the extremely strong anisotropy of $\Delta \varepsilon({\bf k})$,
that could be inferred from angle-resolved photoemission experiments, 
is a clear fingerprint of the anisotropic Rashba effect discussed in this work. 

In summary, we constructed a  $\mathbf{k\cdot p}$ perturbation theory for 
surface states in the presence of SO coupling, and 
derived a generalized  Rashba Hamiltonian for (nearly free) 
electrons on metal  surfaces. We found that in case of 
$C_{2v}$ symmetry, the Rashba interaction gets an 
anisotropic part in first order of \textbf{k}, 
which for Au(110) is found to dominate over the additional, 
well-known symmetric term. The anisotropic 
Rashba term appears due to the mixing of the surface state 
with the bulk states for finite momenta. Even for surfaces 
 of higher symmetry,  this mechanism 
(i.e., the corresponding term in $\bk\cdot\bp$ perturbation theory)
gives a large contribution to the isotropic part of the Rashba Hamiltonian. 
Based on fully relativistic
first-principles electronic structure calculations, we also demonstrated
that a strongly anisotropic Rashba coupling should be 
experimentally observable for  Au(110) surfaces.

Financial support was provided by the Hungarian Research
Foundation (contract no. OTKA K68312, K77771, K73361 and F68726).

\bibliographystyle{myst}


\end{document}